\documentclass[a4paper,12pt]{article} 
\usepackage[centertags]{amsmath}
\usepackage{amsmath}
\usepackage[british]{babel} 
\usepackage{amsfonts}
\usepackage{amssymb}
\usepackage{amsthm}
\usepackage{newlfont}
\usepackage{color}
\usepackage{subfig}
 \usepackage{bbm}

\usepackage{fancyhdr}
\usepackage{graphicx}
\usepackage{fancybox}
\usepackage{geometry}
 \usepackage{psfrag}
 
 \usepackage{multirow}

\usepackage[utf8]{inputenc}
\usepackage{authblk}

\theoremstyle{plain}
  \newtheorem{theorem}{Theorem}[section]

  \newtheorem{lemma}[theorem]{Lemma}
  
\theoremstyle{definition}

\theoremstyle{remark}

\numberwithin{equation}{section}

\DeclareMathOperator{\Tr}{Tr}

\renewcommand{\Re}{\mathrm{Re}\, }

\newcommand\otimesal{\mathop{\hbox{\raise 1.6 ex
  \hbox{$\scriptscriptstyle\mathrm{al}$}
\kern -0.92 em \hbox{$\otimes$}}}}
\newcommand\oplusal{\mathop{\hbox{\raise 1.6 ex
  \hbox{$\scriptscriptstyle\mathrm{al}$}
\kern -0.92 em \hbox{$\oplus$}}}}
\newcommand\Gammal{\hbox{\raise 1.7 ex
\hbox{$\scriptscriptstyle\mathrm{al}$}\kern -0.50 em $\Gamma$}}
\renewcommand\i{\mathrm{i}}

  \let\La=\Lambda

\newcommand{\caA}{{\mathcal A}}
\newcommand{\caB}{{\mathcal B}}

\newcommand{\caD}{{\mathcal D}}

\newcommand{\caF}{{\mathcal F}}

\newcommand{\caH}{{\mathcal H}}

\newcommand{\caL}{{\mathcal L}}

\newcommand{\caV}{{\mathcal V}}

\newcommand{\bbC}{{\mathbb C}}

\newcommand{\bbZ}{{\mathbb Z}}

\newcommand{\opunit}{\text{1}\kern-0.22em\text{l}}
\newcommand{\funit}{\mathbf{1}}

\newcommand{\bsS}{{\boldsymbol S}}

\newcommand{\e}{{\mathrm e}}

\renewcommand{\d}{{\mathrm d}}

\newcommand{\beq}{ \begin{equation} }
\newcommand{\eeq}{ \end{equation} }
\newcommand{\bet}{ \begin{theorem} }
\newcommand{\eet}{ \end{theorem} }

\newcommand{\baq}{\begin{eqnarray}}
\newcommand{\eaq}{\end{eqnarray}}

 \newcounter{smallarabics}
\newenvironment{arabicenumerate}
{\begin{list}{{\normalfont\textrm{\arabic{smallarabics})}}}
  {\usecounter{smallarabics}\setlength{\itemindent}{0cm}
  \setlength{\leftmargin}{5ex}\setlength{\labelwidth}{4ex}
  \setlength{\topsep}{0.75\parsep}\setlength{\partopsep}{0ex}
   \setlength{\itemsep}{0ex}}}
{\end{list}}

\newcounter{smallroman}

\newcommand{\ben}{\begin{arabicenumerate}}
\newcommand{\een}{\end{arabicenumerate}}

\newcommand{\adjoint}{\mathrm{ad}}
\newcommand{\ad}{\adjoint}

\newcommand{\smi}{{\scriptscriptstyle -}}
\newcommand{\spl}{{\scriptscriptstyle +}}
 
\begin{document}

\title{An exponentially local spectral flow for \\possibly non-self-adjoint\\ perturbations of non-interacting quantum spins, \\
inspired by KAM theory}

\author{Wojciech De Roeck and Marius Sch\"utz} 
\affil{{Instituut voor Theoretische Fysica, KU Leuven, Belgium}}

\renewcommand\Authands{ and }
\maketitle

\begin{abstract}
Since its introduction by Hastings in \cite{Has}, the technique of quasi-adiabatic continuation has become a central tool in the discussion and classification of ground state phases. It connects the ground states of self-adjoint Hamiltonians in the same phase by a unitary quasi-local transformation.
This paper takes a step towards extending this result to non-self adjoint perturbations, though, for technical reason, we restrict ourselves here to weak perturbations of non-interacting spins. 
The extension to non-self adjoint perturbation is important for potential applications to Glauber dynamics (and its quantum analogues).    In contrast to the standard quasi-adiabatic transformation, the transformation constructed here is exponentially local. 
Our scheme is inspired by KAM theory, with frustration-free operators playing the role of integrable Hamiltonians.\footnote{some keywords: quantum spin systems, gapped ground states, stationary states of quantum Markov dynamics, frustration-free systems, non-self-adjoint perturbation theory.} \\

\end{abstract}

\vspace{\stretch{1}}

\section{Introduction}
This paper is primarily inspired by the work \cite{BraHas,MicZwo} on ``stability of frustration-free Hamiltonians'' and the results on the quasi-adiabatic continuation \cite{Has,Bachmann}.  In this setting, stability means that when adding a small perturbation to the gapped frustration-free Hamiltonian, one does not change its properties too heavily.  In particular, the gap remains open and the ground state of the perturbed Hamiltonian can be obtained from the original one by applying a quasi-local transformation. 

One of the ultimate aims of our work is to extend such statements to non-self adjoint operators, so as to cover gapped stochastic generators (Glauber dynamics) and their quantum counterparts, see \cite{CubLucMicPer, CraDerSch} for more  background on such a program. At the same time, we try to construct an exponentially local unitary dressing transformation (or `spectral flow') mapping that perturbed ground state to the unperturbed one. This is an improvement over existing results which are restricted to sub-exponentially local transformations. 
In the present paper, these aims will however be only partially achieved, as we will restrict ourselves to a vicinity of non-interacting spins.

Our approach is a KAM procedure that progressively eliminates non-frustration free terms in the Hamiltonian.  We will now give an outline of our approach and simultaneously explain what `frustration-free' here means.  The strategy is restricted to ground states in the `trivial phase' (connected to product states) but it is not restricted per se to states that are only \emph{small} perturbations of product states. As such, the following outline already hints at potential extensions of the method.\\

\noindent \textbf{Acknowledgements.} We wish to thank Nick Crawford for many helpful discussions. Furthermore, we are thankful to the DFG (German Science Foundation) and the Belgian Interuniversity Attraction Pole (P07/18 Dygest) for financial support.

\subsection{Outline}

Let us explain this procedure in an informal way and in the most basic setup for systems of ${s=\frac{1}{2}}$ spins on the lattice. For every finite volume $\Lambda\subset \bbZ^\nu$, the Hilbert space is $\caH={\boldsymbol \otimes}_{x\in\Lambda} \caH_x$, where each $\caH_x\cong \bbC^2$ is the Hilbert space for such a spin at site $x$ with some preferred basis $\lvert\downarrow\rangle, \lvert\uparrow\rangle$. We take 
$\Omega={\boldsymbol\otimes}_{x\in \La} \lvert \downarrow  \rangle_x$ to be a reference state. Throughout the outline, $\Omega$ will appear as an eigenstate of operators for a non-degenerate and gapped eigenvalue $0$ (uniform in the volume). Even though these operators might not be self-adjoint when we are dealing with perturbations that are not self-adjoint, we will speak about $\Omega$ as ground state. Let us also have the operators 
$$
 \lvert \uparrow  \rangle= \sigma^{+}   \lvert \downarrow  \rangle, \qquad    \lvert \downarrow  \rangle= \sigma^{-}   \lvert \uparrow  \rangle.
$$
A frustration-free operator $F$  with ground state $\Omega$ is one that can be written as a sum of local terms $F=\sum_S F_S$, $F_S$ acting in $S\subset \Lambda$, such that $F_S^{} \Omega=F_S^*\Omega=0$.
We are given an operator of the form 
$$
F+V,
$$
where $F$ is frustration free with ground state $\Omega$ and a gap above that ground state, 
and $V$ has only non-frustration-free terms. 
This means that we can write $V=V^++V^-$ with
$$
V^{\pm}= \sum_S  v^{\pm}_S \sigma^{\pm}_S, \qquad   \sigma^{\pm}_S= \prod_{x \in S} \sigma_x^{\pm}.
$$
We moreover assume that $V$ is small, in the sense that all $v^{\pm}_S$ are small (and of course decaying appropriately as $S$ grows large). 
The game consists now of finding an operator $A$ such that 
$$
\e^{\i A} (F+V)  \e^{-\i A} = F'+V'  +\text{a constant},
$$
where $F',V'$ are like $F,V$ above but with $V'$ smaller than $V$.   With foresight, we declare $A$ to be of order $V$ and we expand in orders of $V$:
\begin{equation}\label{eq: higher orders}
\e^{\i A} (F+V)  \e^{-\i A} =F+  \i [A,F] + V +   \text{higher orders in $V$}.
\end{equation}
We now seek $A=A^+ +A^-$ of the same type as $V$ above and so that it satisfies
$$
 \i [A,F] + V =  \text{frustration-free $+$ a constant}.
$$
{A little thought shows that it suffices to solve the equations}
$$
- \i [A^+,F]\Omega =  V^+ \Omega, \qquad    - (\i [A^-,F])^*\Omega =  (V^-)^* \Omega.
$$
Let us focus on the first equation, the second being very similar. 
First, due to the frustration-free property of $F$ it is equivalent to  $\i F A^+\Omega =  V^+ \Omega$ and this is solved by 
\begin{equation}  \label{intro: expression aplus}
A^+\Omega =   -\i \frac{1}{F}V^+ \Omega
\end{equation}
Note that the right-hand side is well-defined, since by the gap assumption $F$ is invertible on the orthogonal complement of $\Omega$, which is also the range of $V^+$. 
The expression \eqref{intro: expression aplus} is indeed an unambiguous prescription for $A^+$ as the latter can only contain $\sigma^+$ operators. Writing $A^+=\sum_S a^+_S  \sigma^+_S$, we get 
$$
a^+_S =  \i    \langle  \sigma^+_S \Omega,  \frac{1}{F}V^+ \Omega \rangle
$$
At this point, one should argue that $a_S^+$ decays sufficiently fast as diameter and/or size of $S$ increases and that it inherits the smallness of $V^+$. This problem is only easy in the case where $F$ is itself a small perturbation of independent spins, and therefore we restrict in this paper to that case. 
Once this step is accomplished, the story writes itself: The new $V'$ is roughly of the order of $(V)^2$, since the leading contribution to the `higher orders' in \eqref{eq: higher orders} is 
$$
-[A,[A,F]] +\i [A,V].
$$
We can hence hope to iterate this procedure to eliminate $V$ entirely. 
One issue that has not yet been addressed is that the new frustration-free operator $F'$, obtained by adding perturbative frustration-free terms to $F$, needs to have a gap to continue the iteration. This likely requires additional assumptions on $F$, but in the case where $F$ is non-interacting, it follows immediately.

\subsection{Comparison with related work}

{There has been related work on (the stability) of frustration-free systems and earlier work on weakly interacting spin systems. 

In \cite{BraHas,MicZwo} the authors prove that the ground state gap for a class of frustration-free Hamiltonians is stable under arbitrary perturbations in the interaction. They use Hastings' spectral flow technique (or quasi-adiabatic continuation) to map the perturbed Hamiltonian by a similarity transformation to a Hamiltonian that is frustration-free (called `locally block diagonal' in \cite{BraHas}) with respect to the unperturbed ground state and for which therefore a gap can be proved more easily. In our work we show that perturbations, which however need not to be self-adjoint, of classical Hamiltonians are similar to frustration-free ones. Moreover, if the perturbation is exponentially quasi-local, so is the new Hamiltonian. This is a new result that cannot be obtained through the sub-exponentially quasi-local spectral flow.

Frustration-freeness was as a helpful property in many studies of gapped systems, see e.g.\ \cite{Nac, SpiSta} on lower bounds of ground state gaps.  It is thus good news that frustration-free systems appear to be rather generic. In \cite{Has2}, Hastings showed that every gapped local Hamiltonian can be rewritten as approximately frustration-free Hamiltonian upon increasing the range of the interaction (the error vanishes in the limit of infinite range). 
As another example, matrix product states in one-dimensional spin chains always possess a frustration-free parent Hamiltonian \cite{FanNacWer}.
A close connection between such parent Hamiltonians and perturbations of classical systems was worked out in \cite{SzeWol}. 

Besides the restriction to frustration-free systems the result \cite{BraHas,MicZwo} rests on assumptions concerning the presence of a local gap and topological order in the unperturbed ground state subspace, which are trivially satisfied in our setting of independent spins and unique ground state.
}

Systems of independent quantum spins, each with a uniformly gapped non-degenerate ground state, are deep in the unique ground state phase regime. Though no longer a product, the ground state remains gapped when adding a weak quasi-local interaction \cite{Yar04} and the infinite volume ground state is unique for arbitrary boundary conditions \cite{Yar05}. 
Various methods have been devised to obtain information on the ground state in such a setting, see for example \cite{KenTas, Yar06} for low-temperature expansions. Another method going back to \cite{DatKen, KirTho} extracts a quasi-local dressing transformation $U$ connecting the ground state $\Omega'$ with the unperturbed product ground state $\Omega$ through $U \Omega =\Omega'$ from certain fixed-point equations. Yarotsky constructed such a dressing transformation for the setting at hand in \cite{Yar04} which is the exponential of an exponentially quasi-local operator (meaning that it can be written as sum of truly local terms whose strength decay exponentially in the size of their support). The transformation is clearly invertible, but not unitary. As mentioned before, by Hasting's technique of quasi-adiabatic continuation \cite{Has,Bachmann}, unitary dressing transformations can be obtained in great generality for the ground states of every two local Hamiltonians that admit a gapped path between them. However, this technique only gives sub-exponential locality of the transformation and it is an interesting open question, whether at least in some cases the locality can be improved, see also \cite{DerSch}. Furthermore, it relies on the spectral calculus for self-adjoint operators, ruling out possible applications for steady states in non-equilibrium systems, which we address in section \ref{sec: SteadyState}.

\section{Setup and Result}
\textbf{Quantum spin system.} In this paper, we consider quantum spin systems on the lattice $\bbZ^\nu$, $\nu \geq 1$. For every finite volume $\Lambda\subset \bbZ^\nu$, its Hilbert space is given by $\caH_\Lambda={\boldsymbol \otimes}_{x\in\Lambda} \caH_x$, where each $\caH_x\cong \bbC^D$ is the Hilbert space describing finite spin degrees of freedom located at site $x$. The bounded operators on $ \caH_\Lambda$, the observables, are denoted by $\caB_\Lambda$ and $\lVert \cdot \rVert$ stands for the operator norm. We will henceforth drop the dependence on volume $\Lambda$, since all our results hold for any $\Lambda$, and all used constants can be chosen uniform in $\Lambda$. As usual, we often identify operators $O\in\caB_{\Lambda'}$, $\Lambda'\subset \Lambda$, with local operators in larger volumes $O\otimes \opunit \in \caB_\Lambda$.
Our result concerns perturbations of independent spins with Hamiltonian $H=H_0+H'$, where
\begin{equation}
H_0 := \sum_{x\in\Lambda} h_x, \quad h_x^{}=h_x^*\in\caB_{\{x\}}.
\end{equation}
We assume that the ground state energy of each single-site contribution $h_x$ is equal to $0$, non-degenerate, and uniformly gapped with gap $g>0$ away from the excited spectrum. $H_0$ is diagonal for a basis consisting of product vectors.
The perturbation is a general operator $H'\in\caB$ and, unlike $H_0$, it does not need to be self-adjoint.  This allows to apply our results to (quantum) Markov generators. 
In fact, within the context of generators,  the restriction of $H_0$ being self-adjoint can also be relaxed and replaced by a condition on the relaxation behaviour, see section \ref{sec: SteadyState} and \cite{DerCraSch} for the setup and generalization we have in mind). This is important in view of applications to non-equilibrium  (non-detailed balance) generators.

\noindent\textbf{Quasi-locality.} Let $\Omega_x\in\caH_x$ be the ground state (vector) of $h_x$, which is unique up to a phase factor, and let hence $\Omega:={\boldsymbol \otimes}_{x\in\Lambda}\Omega_x$ be the ground state of $H_0$. The orthogonal ground state projection at each site $x$ is denoted by $P_x$. We also use the notation $\bar P_x=(\opunit-P_x)$ and, for subsets $S\subset \Lambda$, we abbreviate $P_S= {\boldsymbol \otimes}_{x\in S} P_x$ and $\bar P_S={\boldsymbol \otimes}_{x\in S} \bar P_x$. At each site, the space of observables can be split into the direct sum
\begin{align}\label{eq: DecompSingSite}
& \caB_x = \caB_x^\spl \oplus \caB_x^\smi \oplus \caB_x^n \oplus\mathrm{span}(\opunit),\quad \text{where}\\
& \caB_x^\spl:=\bar P_x \caB_x P_x,\quad  \caB_x^\smi:=P_x \caB_x \bar P_x, \quad \text{and}\quad \caB_x^n:= \bar P_x \caB_x \bar P_x,\nonumber
\end{align}
and, for subsets $S\subset \Lambda$, we also set
\begin{equation*}
\caB_S^\spl := {\boldsymbol \otimes}_{x\in S} \caB_x^\spl, \quad \caB_S^\smi := {\boldsymbol \otimes}_{x\in S} \caB_x^\smi, \quad \text{and} \quad
\caB_S^n := {\boldsymbol \otimes}_{x\in S} \caB_x^n.
\end{equation*}
The operators from the first two spaces may be viewed as those which can locally create excitations out of the ground state or annihilate them at sites within $S$. We use the convention that, if $S=\emptyset$, the above spaces consist only of multiples of the identity $\opunit$. Given the decomposition of single-site observables in \eqref{eq: DecompSingSite}, we can expand every operator $O\in\caB$ accordingly in the form
\begin{align}\label{eq: Decomp}
& O=\sum_{\bsS} O_{\bsS}, \quad O_{\bsS}\in\caB_{\bsS}:= \caB_{S^\spl}^\spl\otimes \caB_{S^\smi}^\smi \otimes \caB_{S^n}^n, \\
& \bsS=(S^\spl,S^\smi, S^n), \quad S^\spl,S^\smi, S^n\subset \Lambda\; \text{ mutually disjoint}.\nonumber
\end{align}
The support of each operator $O_\bsS$, i.e.~the collection of sites where it acts on non-trivially, is denoted by 
\begin{equation*}
S=S(\bsS):=S^\smi\cup S^\spl\cup S^n.
\end{equation*}
\noindent \textbf{A volume intensive norm.} Next, we want to {define a norm} $\lVert\cdot\rVert_\mu$ on $O\in\caB$ which can capture exponential locality of operators and which, unlike the operator norm, does not grow with the volume if $O$ is exponentially local, i.e.~a sum of local operators whose strength decays exponentially in the support. We introduce the function $w(S):=\lvert S \rvert+\lvert S \rvert_c$ on subsets $S\subset \Lambda$, where $\lvert S \rvert$ is the cardinality of $S$ and $\lvert S \rvert_c$ the minimal cardinality of a connected set in $\bbZ^\nu$ containing $S$. Given $\mu \geq 0$, we define the $\mu$-norm on operators $O\in\caB$ as
\begin{equation}
\lVert O \rVert_\mu := \frac{\lVert O_{\boldsymbol \emptyset}\rVert}{\lvert \Lambda \rvert} +\sup_{x\in\Lambda}\sum_{\bsS:\, x\in S} \e^{\mu w(S)} \lVert O_\bsS \rVert.
\end{equation} 
We assign the symbol $\lVert O\rVert_\mu'$ to the above expression without the first term involving $O_{\boldsymbol \emptyset}$. For each operator $O$, we obtained a unique decomposition into a sum of local terms through \eqref{eq: Decomp}.  This decomposition is marginally different from the more conventional (in mathematical physics) representation through \emph{potentials}. More concretely, if 
 $O=\sum_{S\subset \Lambda} O(S)$ for some collection of local operators $O(S)\in\caB_S$, then 
\begin{equation*}
\lVert O\rVert_\mu \leq \sup_{x\in \Lambda} \sum_{S\ni x}4^{\lvert S\rvert}\e^{\mu w(S)}\lVert O(S)\rVert.
\end{equation*}
\textbf{Frustration-freeness.} We say that an operator $O\in\caB$ is {frustration-free} (with respect to the product state $\Omega$) if 
\begin{equation*}
O_\bsS P = P O_\bsS= 0, \quad \text{for all } \bsS.
\end{equation*} 
We can use the decomposition \eqref{eq: Decomp} to characterize frustration-free operators. The terms which can spoil this property are the term $O_{\boldsymbol \emptyset}$ proportional to the identity, which we also write as $O_{\boldsymbol \emptyset}=\caD[O]$, and those consisting exclusively of (excitation) creation operators in $\caB_{S}^\spl$ or of annihilation operators in $\caB_{S}^\smi$. If $\bsS=(S,\emptyset,\emptyset)$, $S\neq \emptyset$, we write $O_\bsS=O^\spl_S$, and in the same way we set $O_\bsS=O^\smi_S$  if $\bsS=(\emptyset,S,\emptyset)$. Furthermore, we define
\begin{align*}
& \caV[O]:=\caV^\smi[O] + \caV^\spl [O]=\sum_{S} O^\smi_S + \sum_{S} O^\spl_S,
\end{align*}
which we call the non-diagonal part of $O$. The frustration-free part of $O$ is then defined as 
\begin{equation*}
\caF[O]:=O-\caD[O]-\caV[O].
\end{equation*}

We now present our main result  on the dressing transformation or spectral flow for the perturbed Hamiltonian
\begin{equation*}
H=H_0+H', \qquad H_0=\sum_{x\in\Lambda} h_x.
\end{equation*}
Since each $h_x\in\caB_x^n$ we have $\lVert H_0\rVert_\kappa=\e^\kappa \sup_x \lVert h_x\rVert$, which appears in the following theorem.

\begin{theorem}[dressing transformation]\label{thm: 1}
Let $\kappa'>\kappa\geq \log 2$ and $\kappa'-\kappa\leq 1$. There is a constant $\epsilon>0$ independent of the volume $\Lambda$, so that, if
\begin{equation}\label{eq: CondThm}
\lVert H'\rVert_{\kappa'}\leq \epsilon\cdot \frac{g^2(\kappa'-\kappa)^2}{\lVert H_0\rVert_{\kappa'}},
\end{equation}
then there are $A_n\in \caB$, $n\geq 1$, so that the following holds true:\\ These operators are summable,
\begin{equation}\label{eq: SumA}
\sum_n \lVert A_n\rVert_\kappa\leq C\cdot \frac {\lVert H' \rVert_{\kappa'}}{g},
\end{equation}
which also shows that the operator $U:=\lim_{n\rightarrow \infty}\e^{-\i A_1}\dots \e^{-\i A_n}$ exists in every finite volume $\Lambda$. It defines a similarity transformation,
\begin{equation}\label{eq: TrfHam}
U^{-1} (H_0+H')U^{} :=H_F= H_0+d\cdot {\textnormal \opunit} + F,
\end{equation}
where $d\in\bbC$ and $F$ is frustration-free with respect to $\lvert \Omega \rangle$ and bounded as 
\begin{equation}\label{eq: BoundFThm}
\lVert F\rVert_{\kappa}\leq C \cdot \frac{\lVert H_0\rVert_{\kappa'}\lVert H'\rVert_{\kappa'}}{g(\kappa'-\kappa)}.
\end{equation}
The constant $C$ does not depend on the volume.
$H_0 +H'$ and $H_F$ have the same spectrum and $d$ is a non-degenerate eigenvalue gapped away from the rest of the spectrum. If the perturbation $H'$ is self-adjoint, then all the $A_n$ can be chosen to be self-adjoint. In this case, $ \Omega $ is the non-degenerate ground state of $H_F$, the operator $U$ is unitary and $\Omega'=U \Omega $ is the non-degenerate ground state of $H_0+H'$.
\end{theorem}

\noindent\textbf{Locality of the dressing transformation.}
If $H'$ is self-adjoint, we may view $U$ as the unitary evolution operator $U(t)$, $t\in[0,1]$, arising from a time-dependent piecewise constant Hamiltonian $A(t)$.
In this setting, Lieb--Robinson bounds yield quasi-locality for the Heisenberg dynamics of observables, $\alpha_t(O)=U^*_t O U_t$, $O\in\caB$:  Let $O$ be a local observable with support in $S$, and set $S_r:=\{x\,;\,d(x,S)\leq r\}$ for the support extended up to range $r\geq 0$. Then there is an observable $O_r$ with support in $S_r$, so that
\begin{equation}\label{eq: LocFromLR}
\lVert U^* O U -O_r \rVert \leq C\cdot \lvert S\rvert \lVert O\rVert (\e^{\kappa vt}-1)\e^{-\kappa \cdot r}
\end{equation}
for time $t=1$ and Lieb--Robinson velocity $v$. Again $C\geq 0$ is a constant which of course does not depend on the volume nor on the particular observable $O$.
To obtain the above bound, we {define the piecewise constant time-dependent Hamiltonian $A(t)$, $t\in[0,1]$, through $A(0)=A_1$ and}
\begin{equation*}
A(t)=\frac{\textstyle \sum_{k=1}^\infty \lVert A_k \rVert_\kappa}{\lVert A_n\rVert_\kappa }\, A_n \quad \text{if }\quad\sum_{k=1}^{n-1} \lVert A_k \rVert_\kappa < t\cdot \sum_{k=1}^\infty \lVert A_k \rVert_\kappa\leq \sum_{k=1}^{n} \lVert A_k \rVert_\kappa.
\end{equation*}
{By construction the $\kappa$-norm of the Hamiltonian $A(t)$ is bounded  by $\sum_{k=1}^\infty \lVert A_k \rVert_\kappa$ for all times $t\in[0,1]$}, so that we can find a uniform in $t$ Lieb--Robinson velocity $v\sim \lVert H'\rVert_{\kappa'}/(\kappa g)$ from \eqref{eq: SumA}. 
{See for example ref.~\cite{Bachmann} together with \cite{NacOgaSim} for a version of Lieb--Robinson bounds for time-dependent Hamiltonians. There they assume strongly continuous time dependence, but the proof remains valid in our setting. (Note that $A(t)$ is piece-wise constant on a finite number of intervals up to times $1-\delta$ for arbitrary $\delta>0$. This parameter can be chosen smaller for increasing volumes to bound the effect of the remaining time evolution uniformly in the volume.)}

With this locality result, the problem of computing expectation values of local observables for the ground state of $H_0+H'$ can therefore be transferred to computing those of quasi-local observables (with exponential decay) with respect to the simple product state $ \Omega$.

If the perturbation $H'$ and hence possibly the $A_n$ are not self-adjoint, then the above arguments involving a unitary time evolution are no longer valid, but we can still prove a locality property in our specific setting that is very similar (in some direction even stronger) to \eqref{eq: LocFromLR}. Let the observable $O$ be exponentially localized, say around a certain site $x\in\Lambda$, and we want to express that $U^{-1}OU$ remains localized in that way.   For that purpose, we introduce the norm
\begin{equation*}
\lVert O \rVert_{\mu,x} :=\lVert O_{\boldsymbol{\emptyset}}\rVert +\sum_{\bsS}\e^{\mu w_x(S)}\lVert O_{\bsS} \rVert,\qquad w_x(S):=w(S\cup \{x\}),
\end{equation*}
where $\mu\geq 0$ again indicates an exponential decay rate.
\begin{theorem}[locality of dressing transformation]\label{thm: Thm2}
In the same setting and under the same condition \eqref{eq: CondThm} of Theorem \ref{thm: 1}, we have
\begin{align*}
&\lVert U^{-1} O U \rVert_{\kappa,x}\leq C\cdot \lVert O \rVert_{\kappa',x},\qquad
\lVert U^{-1} O U \rVert_{\kappa}\leq C\cdot \lVert O \rVert_{\kappa'},
\end{align*}
for all $O\in\caB$ and $x \in \Lambda$, where $C\geq 0$ is a constant independent of the volume.
\end{theorem}
Note that this result gives locality both of local observables around $x$, as of observables that are sums of local terms throughout the whole volume. 

As opposed to \eqref{eq: LocFromLR}, this result shows that the local terms of the transformed operator $U^{-1} O U$ decay exponentially in the size of their support and not only in the distance to the support of the original operator $O$. On the other hand, when reverting to the formulation of \eqref{eq: LocFromLR}, there is roughly speaking a prefactor that grows exponentially $\sim \e^{(\kappa'-\kappa)w(S)}$ in the support of $O$ instead of only linearly $\sim \lvert S \rvert$ as in \eqref{eq: LocFromLR}.

\subsection{Stationary states of (quantum) Markov Dynamics} \label{sec: SteadyState}

The dressing transformation introduced above cannot only be applied to ground states but also to stationary states in weakly coupled (quantum) spin systems with Markovian time evolution. Such dynamics are often used to effectively describe or model open quantum systems under the influence of dissipation and possibly non-equilibrium driving. 
{In  quantum computation, one harbours the hope to prepare quantum states as stationary states from engineered quantum dissipative dynamics \cite{KraEtAl,VerWolCir}. The stability of these systems and in particular of the corresponding stationary states under perturbations is obviously important for such ideas. Since there is a well-established notion of stability and of phases for gapped ground states (allowing for a spectral flow between states of the same phase: a strong manifestation of stability), one may wonder if some aspects can be transferred to the study of stationary states in dynamics from quasi-local gapped generators (Lindbladians). Whereas locality results analogous to Lieb--Robinson bounds in Hamiltonian systems were obtained for quantum Markov dynamics \cite{Pou,NacVerZag}, Hastings' spectral flow technique cannot be generalized to non-normal operators and operators whose spectrum is not confined to the reals (both is possible for general generators) in a straight forward way. In this case it is not clear whether persistence of a gap is sufficient for stability. The work of \cite{CubLucMicPer,BraCubLucMicPer} concentrates on the stability of the dynamics and stationary states of rapidly mixing systems, which involves a condition on the (fast) relaxation behavior of the dynamics that is stronger than a gap. In \cite{KasTem} the authors discuss the connection between gaps, relaxation times, and Log-Sobolev inequalities for general quantum Markov dynamics. See also \cite{Zni} for related work on extended quasi-local systems.}

We will first briefly outlay the setup and state the result for quantum Markovian dynamics. {Our theorem applies to weakly coupled systems in the uniqueness regime as introduced in \cite{CraDerSch}.} Markov jump processes for classical spin systems will be discussed separately, even though they are in fact a special case of the quantum formalism.

We still work on a quantum lattice system with the same Hilbert space as defined in the beginning of this section, but, concerning the transfer of the result, the role of $\caH_x$ will be played by the space of single site observables $\caA_x:=\caB(\caH_x)$ and the role of $\caH_\Lambda$ is consequently taken over by $\caA_\Lambda:={\boldsymbol \otimes}_{x\in\Lambda}\caA_x$. Most naturally, these spaces should be endowed with the operator norm to form a $C^*$ algebra, but we furnish them with Hilbert space structure to allow a more direct application of our result. Towards the end of this discussion we get back to this issue. For each site $x\in\Lambda$, let $\rho_x$ be a density matrix on $\caH_x$ with full rank, e.g.~a thermal state $\rho_x\sim \e^{-\beta_x V_x}$ for some inverse temperature $\beta_x$ and Hamiltonian $V_x$. Then we define an inner product on $\caA_x$ through
\begin{equation*}
\langle A,B\rangle_{\rho_x}=\Tr_{\caH_x} (\rho_x A^* B)=\rho_x(A^* B), \qquad A,B \in\caA_x,
\end{equation*}
where we slightly abused notation in that we use the same symbol for density matrices and the associated state (functional).
In analogy to the product ground state $\Omega$ we are here dealing with a product stationary state $\rho:={\boldsymbol \otimes_{x\in\Lambda}\rho_x}$. We define $\caA$ as the tensor product of Hilbert spaces $\caA_x$ with inner product denoted by $\langle \cdot,\cdot\rangle_\rho$. In place of each single site Hamiltonian $h_x$, we consider a Lindblad generator $l_x\in \caB(\caA_x)$. It generates a quantum Markov semigroup of completely positive and identity preserving (super-) operators $\e^{t\,l_x}$, $t\geq 1$, which define the time evolution $A(t)=\e^{t\,l_x}A$ of observables. Such generators annihilate the identity $\opunit$ and cannot have eigenvalues with positive real part. To be in line with the requirements of our theorem, we assume that each generator $l_x$ is self-adjoint and that $0$ is a non-degenerate and uniformly gapped eigenvalue, which implies that at each site the dynamics is relaxing to the (unique) stationary state $\tau(A_t)\rightarrow\rho_x(A)$,  $t\rightarrow \infty$, for all observables $A$ and initial states $\tau$. As mentioned before, the assumption of self-adjointness should not be necessary in this setting and replaceable by assuming a spatially uniform relaxation rate, but we do not give the details of the proof (more concretely, Lemma \ref{lem: Generator} can be adapted following \cite{CraDerSch}). We set
\begin{equation} \label{eq: Lindblad}
L_0 := \sum_{x\in\Lambda} l_x,\quad\text{and}\quad L = L_0 + L', 
\end{equation}
where the perturbation $L'$ is an arbitrary Lindblad generator. If it is exponentially local and weak enough in the sense that $\lVert L' \rVert_{\kappa'}$ is small, then as a result of the theorem there is a dressing transformation,
\begin{equation}\label{eq: TrfLindblad}
U^{-1} (L_0+L') U = L_0 + F,
\end{equation}
where $F$ annihilates the identity as frustration free operator. By definition $0$ is an eigenvalue of the left hand side, and therefore the constant $d$ as appears in \eqref{eq: TrfHam} does not show up. To make the meaning of our definition of frustration-freeness more clear in this context (the terminology may not be too suitable here) note that $P$ is the rank one projection taking observables $A$ to $\rho(A)\opunit$. Therefore $F$ satisfies 
\begin{equation*}
\rho(F A)=\rho(PFA)=0.
\end{equation*}
By the gap stability claimed in Theorem \ref{thm: 1}, the perturbed dynamics retains a unique stationary state, which we denote with $\rho'$. Since $\rho$ is the stationary state for the dynamics generated by $L_0$, we have
\begin{align*}
&\rho(L_0 A)=\rho(U^{-1}(L_0+L')U A) =0
\end{align*}
and therefore also
\begin{align*}
&\rho(U^{-1}(L_0+L') A) =0
\end{align*}
for all observables $A$. It implies that the perturbed stationary state is given through
\begin{equation*}
\rho'(A)=\lambda\cdot\rho\bigl(U^{-1} A\bigr),
\end{equation*}
where $\lambda\in\bbC$ is a non-zero normalization constant.
The null space of both $L_0+L'$ and $L_0+F$ is spanned by the identity $\opunit$, and from \eqref{eq: TrfLindblad} we then conclude that $U(\opunit)=\lambda\opunit$. The density matrices are related through
\begin{equation}\label{eq: NonComRN}
\rho^{-1}\rho' = \lambda \cdot \bigl(U^{-1}\bigr)^* \opunit.
\end{equation}
Since $\rho$ is a product, taking the adjoint of operators on $\caA$ does not change their locality, and in particular, we have 
\begin{equation*}
\lVert (U^{-1})^*\rVert_\kappa=\lVert U^{-1}\rVert_\kappa.
\end{equation*}
We can get another type of locality estimate by using Theorem \ref{thm: Thm2} and rewriting expectations with respect to $\rho'$ in the following way. For $A\in \caA$, we denote with $\caL_A \in \caB(\caA)$ the operator defined by left multiplication with $A$, then
\begin{align*}
& \rho'(A)=\rho(U^{-1} \caL_A \opunit)=\rho(A'),\\
&\text{for } A':=U^{-1} \caL_A U \opunit.
\end{align*}
By Theorem \ref{thm: Thm2} we have therefore again obtained a way to express expectation values of local observables for $\rho'$ in terms of the simple product state $\rho$ and an exponentially quasi-local observable. More precisely and having in mind that $A$ and hence also $\caL_A$ may be localized near some site $x$, we find that 
\begin{equation*}
\lVert U^{-1} \caL_A U\rVert_{\kappa,x}\leq C \cdot \lVert\caL_A \rVert_{\kappa,x}
\end{equation*}
and therefore $A'$ can be written as sum of local operators
\begin{align*}
& A' =\sum_{S\subset\Lambda} A'_S, \qquad A_S'\in \caA_S,\\
\text{with decay}\quad& \sum_{S\subset \Lambda} \e^{\kappa w_x(S)} \lVert A_S'\rVert\leq C\cdot\lVert A \rVert_{\kappa',x}. \hphantom{mmmmmmm}
\end{align*}
The norm appearing in the sum is still the weighted Hilbert--Schmidt norm in $\caA$, but obviously we can deduce exponential decay of $A'$ at a smaller rate also for the more natural operator norm,
\begin{equation*}
\bigl\lVert A'_S \bigr\rVert_\mathrm{op}^{2} \leq c^{\lvert S\rvert} \,\lVert A_S'\rVert^2, 
\end{equation*}
with $1/c$ the smallest eigenvalue of $\rho_x, x \in S$. 

\subsection{Classical stochastic processes}
Markov jump processes on classical spins, or interacting particle systems, are embedded in the above formalism. We repeat the discussion for the convenience of the reader. Let $\Sigma_x=\{\sigma_{1}^{\scriptscriptstyle (x)},\dots,\sigma_{D}^{\scriptscriptstyle (x)}\}$ be an orthonormal basis of $\caH_x$ for every site $x\in\Lambda$. We identify each $\Sigma_x$ with the configuration space of a single classical spin and the Cartesian product $\Sigma = {\small \prod}_{x\in\Lambda}\Sigma_x$ is the configuration space of the spin lattice. The observables for each spin are defined as the functions on $\Sigma_x$, which can be identified with $\caD_x$, the subset of operators in $\caA_x$ that are diagonal for the basis $\Sigma_x$. Again, the natural choice of norm for observables would be the $\sup$ norm (corresponding to the operator norm), but we choose a Hilbert space structure to apply our result. Let $\nu_x$ be a strictly positive probability measure on $\Sigma_x$, then we set $\caD_x:=l_2(\Sigma_x,\nu_x)$ and
\begin{equation*}
\caD := {\boldsymbol \otimes}_{x\in\Lambda} l_2(\Sigma_x,\nu_x), \cong l_2 (\Sigma,\nu)
\end{equation*}
where $\nu$ is the product of the measures $\nu_x$. We assume that $\nu_x$ is the unique stationary measure of a Markov jump process with self-adjoint and gapped generator $l_x$ (corresponding to a Lindbladian acting non-trivially only on the subspace $\caD_x\subset\caA_x$). Given any Markov generator $L'$ on $\caD$, whose norm $\lVert L'\rVert_{\kappa}$ is small enough, we obtain a generator $L$ of weakly coupled classical spins just as above in \eqref{eq: Lindblad} with unique stationary measure denoted by $\nu'$. Using the Theorems in the same way as above, we find that, for all $f\in \caD$ (thinking about a  local function near some site $x$),
\begin{align*}
\nu'(f)=\nu(f') \quad\text{with}&\quad f':= U^{-1} m_f U \funit = \sum_{S\subset\Lambda} f'_S, \\
\text{and}&\quad \sum_{S\subset \Lambda} \e^{\kappa w_x(S)} \lVert f_S'\rVert\leq C\cdot\lVert f \rVert_{\kappa',x},
\end{align*}
where $m_f$ is the operator multiplying by $f$, where $\funit$ is the constant function, and each $f'_S$ is a function depending only on the configuration of spins in a subset $S\subset\Lambda$. Again, we can find an exponential decay estimate for $f'$ also with respect to the sup norm if the rate $\kappa$ could be chosen large enough and if the measures $\nu_x$ are bounded uniformly from below for all $x$.

As a final remark in this section, note that
\begin{equation*}
\nu'\bigl(f\bigr)=\lambda \cdot \nu\bigl(U^{-1}f\bigr)=\lambda \cdot\nu\bigl(f\cdot(U^{-1})^*_{} \funit\bigr)
\end{equation*}
in terms of the adjoint operator of $U^{-1}$ in  $l_2(\Sigma,\nu)$ and a normalization constant $\lambda\in\bbC$. In analogy with \eqref{eq: NonComRN}, we have hence obtained an explicit expression for the Radon--Nikodym derivative
\begin{equation*}
\frac{\d \nu'}{\d \nu}=(U^{-1})^*_{} \funit.
\end{equation*}
In our perturbative setup the exponentials appearing in $U^{-1}=\lim_{n\rightarrow \infty}\e^{\i A_n}\dots \e^{\i A_1}$ should be manageable with high temperature cluster expansion techniques (where high temperature corresponds to the smallness of $\lVert A_n\rVert_\kappa$) to show that the above quotient is a positive function and that it moreover can be written as a the exponential
\begin{equation*}
(U^{-1})^*_{} \funit=\exp(\Phi)
\end{equation*}
of a potential $\Phi$ which is an exponentially local function. See e.g.~references \cite{NetRed, DerMaeNetSch} for the background of this claim.

\section{The Proof \\{\normalsize-- Using an Idea from KAM Theory and \\[-0.2\baselineskip] Properties of Frustration Free Hamiltonians --}}
We now describe our main iteration scheme to transform the Hamiltonian. It is inspired by KAM theory. 
We rename the Hamiltonian $H_1=H_0+H'$ and split it (in a unique way) into a sum,
\begin{equation*}
H_1= H_0+d_1+F_1+V_1,
\end{equation*}
where $d_1=\caD[H']\in \bbC$ is proportional to the identity, $F_1=\caF[H']$ is the frustration-free part, and $V_1=\caV[H']$ is the non-diagonal remainder of the perturbation $H'$. Starting from our Hamiltonian $H_1$, we recursively define a sequence of Hamiltonians $H_n$, $n\geq 1$, where $H_{n+1}$ is obtained from $H_n$ through
\begin{equation}\label{eq: Trf}
H_{n+1}:= \e^{\i \ad_{A_n}}H_n=\e^{\i A_n} H_n \e^{-\i A_n},
\end{equation}
and just below in \eqref{eq: DefG} we will specify $A_n$ as a function of $H_n$, which is self-adjoint if $H_n$ is self-adjoint. In that case, the $H_n$ hence define a sequence of unitarily equivalent Hamiltonians. We denote with $d_n$, $F_n$, and $V_n$ the constant, frustration-free, and non-diagonal part of $H_n-H_0$ respectively. 

\begin{lemma}\label{lem: Gap}
Let $F\in \caB$ be frustration-free and $\lVert F \rVert_{\mu=0}<g/2$, then $0$ is a non-degenerate eigenvalue of $H_0+F$ (for eigenstate $\Omega$) and the real part of the remaining spectrum is larger than $g/2$.
\end{lemma}

\noindent This lemma on the gap stability of $H_0$ upon frustration-free perturbations is proven later in section \ref{sec: Proofs} (as well as all other lemmata). It allows to define the reduced resolvent of $H_0+F_n$ for the eigenvalue $0$, i.e., the operator $R_n$ that satisfies
\begin{equation*}
R_n(H_0+F_n)=(H_0+F_n)R_n=\bar P,
\end{equation*}
at least as long as $\lVert F_n\rVert_{\mu= 0}\leq g/2$. Under this condition, we define the operator $A_n$, which was introduced above, as
\begin{equation}\label{eq: DefG}
A_n:=\i\caV^\smi [V_n R_n]-\i\caV^\spl [R_n V_n].
\end{equation}

\begin{lemma}\label{lem: PropG}
$A_n$ solves the equation $V_n+\i\caV\bigl[[A_n,H_0+F_n]\bigr]=0$
\end{lemma}

\noindent This property of $A_n$ is central in our construction, as it enables us to rewrite the transformation as follows. With this lemma, we see a cancellation of terms when expanding each exponential in
\begin{equation*}
H_{n+1}=d_n+\e^{\i\ad_{A_n}}(H_0+F_n)+\e^{\i\ad_{A_n}}V_n.
\end{equation*}
and we get $H_{n+1}= H_0+d_{n+1}+F_{n+1}+V_{n+1}$ with
\begin{align}\label{eq: RecStep}
& d_{n+1}=d_n+\caD\bigl[\i[A_n,H_0+F_n]+E_{n+1}\bigr],\\
& F_{n+1}=F_n+ \caF\bigl[\i[A_n,H_0+F_n]+E_{n+1}\bigr],\label{eq: RecStepF}\\
& V_{n+1}=\caV[E_{n+1}],
\end{align} 
where we introduced
\begin{equation} \label{eq: DefE}
 E_{n+1}:=\sum_{k=1}^\infty\frac{\ad_{\i A_n}^k }{k!} \biggl(\frac{\i\bigl[A_n,H_0+F_n\bigr]}{k+1}+V_n\biggr).
\end{equation}
The operators $A_n$ are particularly accessible in perturbative expansions, as they are defined through the resolvent of a frustration-free perturbation of $H_0$. As a consequence, we will be able to show that, roughly speaking, the magnitude of $A_n$ is the same as that of $V_n$ as long as $F_n$ remains small enough. We can start the recursion with a small $V_1$. Assuming for a moment that the $V_n$ at least do not grow, $E_{n+1}$ can in this sense be viewed as a second order contribution in $V_n$. The important observation then is that 
$V_{n+1}=\caV[E_{n+1}]$
is quadratic in the precursor $V_n$, which provides the (super-exponential) convergence of the procedure common from KAM. The argument can be closed consistently realizing that the differences $ F_{n+1}-F_{n}$ decay just as fast as $V_n$. In summary we therefore find that $V_n\rightarrow 0$ super-exponentially as $n\rightarrow \infty$ and that the map $\lim_{n\rightarrow \infty} \e^{\ad_{\i A_n}}\dots\e^{\ad_{\i A_1}}$ will take the Hamiltonian $H_0+H'$ to $H_F=\lim_{n\rightarrow \infty} H_n$, which is frustration-free apart from a constant.
We now supply the quantitative estimates.

\subsection{Convergence of the recursion relation}

Unfortunately, we did not manage to set up the whole procedure based on a single norm and we therefore consider a family of norms by fixing a strictly decreasing sequence of decay rates,
\begin{equation*}
\kappa_n = \kappa +(\kappa'-\kappa)/n, \quad n\geq 1,
\end{equation*}
which lie between $\kappa'$ and $\kappa$, so that $\kappa_1=\kappa'$ and $\kappa$ is approached as $n\rightarrow \infty$. Note also that the differences decrease as
\begin{equation*}
\delta \kappa_{n+1}:=\kappa_n-\kappa_{n+1} = \frac{\kappa'-\kappa}{n(n+1)}.
\end{equation*}
We will often abbreviate the associated norms $\lVert \cdot \rVert_{\kappa_n}$ simply by $\lVert \cdot \rVert_n$, and we also introduce the abbreviations 
\begin{align*}
&e_1:= \lVert H' \rVert_{\kappa'}, \quad e_{n}:=\lVert E_{n}\rVert_{2n},\quad n\geq 2,\\
& f_n:=\lVert F_{n}\rVert_{2n}, \quad v_n:=\lVert V_{n}\rVert_{2n}, \quad a_n:=\lVert A_{n}\rVert_{2n}, \quad n\geq 1.
\end{align*}
We chose the $2n$-norm instead of the $n$-norm in our estimates at the $n$-th step only because of later notational convenience. The shifts $d_n$ are mostly irrelevant in the construction, since $A_{n+1}$ and therefore also $E_{n+1}$ does not depend on it. One should not expect volume independent convergence and upper bounds for the $d_n$, which present overall energy renormalizations, but rather for the densities $d_n/\lvert \Lambda \rvert$.

Next, we state the main tools of our proof. The first of the following three lemmata shows that indeed $a_n\sim v_n$ as long as there is a uniform upper bound for the $f_n$. Its proof is based on a perturbative expansion for the resolvent of a frustration-free Hamiltonian of the form $H_0+F$. The other two lemmata give general estimates on the commutator and, basically, the exponential of operators for our particular type of volume intensive norms.

\begin{lemma}\label{lem: Generator}
Let $\mu\geq 0$ and $F,V\in \caB$, where $F=\caF[F]$ is frustration-free with $\lVert F \rVert_{\mu}<g/4$ and $V=\caV[V]$ non-diagonal. The reduced resolvent $R$ of $H_0+F$ for the eigenvalue $0$ exists by Lemma \ref{lem: Gap} and $A=\i\caV^\smi [V R]-\i\caV^\spl [R V]$ is well-defined. It satisfies
\begin{equation}\label{eq: BoundG}
\lVert A\rVert_\mu\leq  \frac{8 \lVert V\rVert_{\mu}}{g/4-\lVert F\rVert_\mu} .
\end{equation}
\end{lemma}

\begin{lemma}\label{lem: Commutator}
Let $\mu'>\mu\geq\log 2$ and $A,B\in \caB$, then 
\begin{equation}
 \bigl\lVert [A,B]\bigr \rVert_\mu \leq \frac{8\lVert A\rVert_{\mu'} \lVert B \rVert_{\mu'}}{\mu'-\mu}.
\end{equation}
\end{lemma}

\begin{lemma}\label{lem: Exponential}
Let $\mu'>\mu\geq\log 2$, $\mu'-\mu\leq 1$, $A\in\caB$ with $\lVert A \rVert_{\mu'}\leq(\mu'-\mu)/6$, and $B^{\scriptscriptstyle (k)}\in \caB$, $k\geq 1$, then
\begin{equation}\label{eq: BoundExp}
\Bigl\lVert \sum_{k=1}^\infty \frac{\ad_A^k B^{\scriptscriptstyle (k)}}{k!}\Bigr\rVert_{\mu}\leq \frac{252 \,b\lVert A\rVert_{\mu'}}{\mu'-\mu},
\end{equation}
where $b=\sup_k \lVert B^{\scriptscriptstyle (k)}\rVert_{\mu'}$.
\end{lemma}

{
We will now show inductively for all $n\geq 1$ that $e_n\leq e_1/n^4$ and $f_n\leq g/8$  if 
\begin{equation*}
\epsilon:= \frac{e_1 \lVert H_0\rVert_{\kappa'}}{g^2(\kappa' -\kappa)^2}
\end{equation*}
is small enough (independent of $n$). The choice of proving a decay $\sim 1/n^4$ is rather arbitrary at this point, and we will later see that it is indeed faster than exponential. The type of decay obviously cannot be detected from a finite number of first terms in the sequence $e_n$ that are relevant for the induction. In the following $c,c'>0$ stand for numerical constants that we do not bother to specify and their value may be different in each expression (independent of $n$). For $n=1$ the claim concerning $e_1$ is trivial and $f_1\leq g/8$ follows from 
\begin{equation*}
f_1 \leq \lVert F_1 \rVert_{\kappa'} \leq e_1 \leq \frac{\epsilon g^2 (\kappa'-\kappa)^2}{\lVert H_0\rVert_{\kappa'}}\leq \epsilon g
\end{equation*}
if $\epsilon \leq 1/8$. For the last inequality we used that $g\leq \lVert H_0\rVert_{\kappa'}$ and the assumption $\kappa'-\kappa \leq 1 $ (both inequalities will be used repeatedly in the following). Assume that the claim holds up to some $n\geq 1$, then, by Lemma \ref{lem: Generator} employed with $\mu=\kappa_{2n}$ and $A=A_n$, we can confirm 
\begin{equation}\label{eq: UseLemmaA}
\lVert A_n\rVert_{2n+1} \leq a_n \leq c\cdot \frac{v_n}{g}\leq c\cdot \frac{e_n}{g}\leq c\cdot \frac{e_1}{n^4g}\leq c\cdot\frac{\epsilon (\kappa' -\kappa)}{n^4}\leq \delta\kappa_{2(n+1)}/6,
\end{equation}
for $\epsilon>0$ small enough. This is the working assumption for using Lemma \ref{lem: Exponential} with  
\begin{equation*}
B^{\scriptscriptstyle (k)}=\i\frac{[A_n,H_0+F_n]}{k+1}+V_n
\end{equation*}
as from \eqref{eq: DefE} and with $\mu'=\kappa_{2n+1}=\mu+\delta\kappa_{2(n+1)}$ to get the upper bound
\begin{align*}
e_{n+1} & \leq c\cdot \frac{\lVert A_n\rVert_{2n+1}}{\delta\kappa_{2(n+1)}} \bigl\lVert [A_n,H_0+F_n] +V_n \bigr\rVert_{2n+1}\\
&\leq c'\cdot \frac{n^2 a_n}{\kappa'-\kappa}\bigl( \bigl\lVert [A_n,H_0+F_n]\bigr\rVert_{2n+1} +e_n \bigr).
\end{align*}
The commutator term can be estimated through Lemma \ref{lem: Commutator} with $\mu'=\kappa_{2n}=\mu+\delta\kappa_{2n+1}$ and again Lemma \ref{lem: Generator} to give
\begin{align}\label{eq: BoundCommutator}
 \bigl\lVert [A_n,H_0+F_n]\bigr\rVert_{2n+1} &\leq c\cdot \frac{a_n }{\delta\kappa_{2n+1}}\bigl(\lVert H_0\rVert_{2n+1} +f_n\bigr)\nonumber\\
 &\leq c'\cdot\frac{n^2 a_n}{\kappa'-\kappa}\lVert H_0\rVert_{\kappa'},
\end{align}
where we used that  $1/\delta \kappa_{2n+1}\leq 2 n^2 /(\kappa'-\kappa)$ and that $f_n\leq g/8\leq\lVert H_0\rVert_{2n+1}\leq \lVert H_0\rVert_{\kappa'}$ by the induction hypothesis. If we insert this inequality in the previous estimate and by bounding $a_n$ in terms of $e_n$ as in \eqref{eq: UseLemmaA} we finally arrive at
\begin{align}\label{eq: SuperExpConv}
e_{n+1}&\leq c\cdot\frac{n^2 e_n}{g(\kappa'-\kappa)}\Bigl(\frac{n^2 e_n \lVert H_0\rVert_{\kappa'}}{g(\kappa'-\kappa)}+e_n\Bigr)\nonumber\\
&\leq  c'\cdot \frac{\epsilon}{e_1} \bigl(n^2 e_n\bigr)^2, 
\end{align}  
which together with $e_n\leq e_1/n^4$ also implies the first part of the induction step, 
\begin{equation}\label{eq: IndStep}
e_{n+1}\leq e_1/(n+1)^4,
\end{equation} for $\epsilon>0$ small enough. With this result we can easily obtain the second part of the induction claim, that $f_{n+1}\leq g/8$, as follows. Recalling the form of $F_n$ in \eqref{eq: RecStepF} we find
\begin{align*}
f_{n+1} &\leq f_n + \bigl\lVert [A_n,H_0+F_n]\bigr\rVert_{2(n+1)}+e_{n+1}\\
&\leq f_n + c\cdot\frac{n^2 a_n\lVert H_0\rVert_{\kappa'}}{\kappa'-\kappa}+e_1/(n+1)^4,
\end{align*}
since the commutator term is bounded by \eqref{eq: BoundCommutator}. Using once again the bound \eqref{eq: UseLemmaA} on $a_n$ together with $g\leq \lVert H_0\rVert_{\kappa'}$ and $\kappa'-\kappa \leq 1$ and \eqref{eq: IndStep}, we furthermore arrive at
\begin{align*}
f_{n+1}&\leq f_n+c \cdot \frac{e_1 \lVert H_0\rVert_{\kappa'}}{g(\kappa'-\kappa)(n+1)^2}=f_n+c\cdot\frac{\epsilon }{(n+1)^2}g(\kappa'-\kappa),
\end{align*}
{for $\epsilon>0$ small enough. A factor $4$ from $n^{-2}\leq 4 (n+1)^{-2}$ was absorbed in the constant $c$.}
 By the same computation we can show
\begin{align*}
\bigl\lVert F_{n+1}-F_n\bigr\rVert_\kappa\leq \bigl\lVert F_{n+1}-F_n\bigr\rVert_{n+1}&\leq \bigl\lVert [A_n,H_0+F_n]\bigr\rVert_{2(n+1)}+e_{n+1}\\
&\leq c\cdot\frac{\epsilon }{(n+1)^2}g(\kappa'-\kappa),
\end{align*}
which of course implies that $F_n\rightarrow F$ in the norm $\lVert \cdot\rVert_\kappa$. Since we have $f_1\leq \epsilon g (\kappa'-\kappa)$, we obtain
\begin{align*}
f_{n+1}\leq c\cdot\epsilon g (\kappa'-\kappa)=c\cdot \frac{\lVert H_0 \rVert_{\kappa'}\lVert H' \rVert_{\kappa'}}{g (\kappa'-\kappa)},
\end{align*}
which not only finishes the induction but also presents the theorem's bound \eqref{eq: BoundFThm} on $\lVert F\rVert_\kappa$. In the same way it follows that the differences of the constants $d_n$ decrease at least as
\begin{align*}
\lvert d_{n+1}-d_n\rvert& \leq \lvert \Lambda\rvert \bigl(\bigl\lVert [A_n,H_0+F_n]\bigr\rVert_{2(n+1)}+e_{n+1}\bigr)\\
& \leq c\cdot\lvert \Lambda\rvert\frac{\epsilon }{(n+1)^2}g(\kappa'-\kappa)
\end{align*}
implying that $d_n/\lvert \Lambda\rvert\rightarrow d/\lvert \Lambda\rvert$ converges uniformly for all volumes $\Lambda$. The non-frustration-free parts $V_n\rightarrow 0$ in $\lVert \cdot\rVert_\kappa$ norm because $v_n\leq e_n$. To finish the proof of Theorem \ref{thm: 1}, we need to show the summability of the $\lVert A_n\rVert_\kappa$ according to \eqref{eq: SumA}, which follows from
\begin{equation*}
\lVert A_n\rVert_\kappa\leq a_n\leq c\cdot\frac{e_n}{g}\leq c \cdot\frac{e_1}{g n^4}\leq c\cdot\frac{ \lVert H' \rVert_{\kappa'}}{g n^4 }.
\end{equation*}
As an aside, note that \eqref{eq: SuperExpConv} in our proof actually shows that the convergence speed of our procedure is faster than $\sim 1/n^4$ and even faster than exponential if $\epsilon>0$ is small enough, meaning for example that $e_{n+1}/e_n\leq n^4\epsilon^n < 1$ for all $n\geq 1$.
}

\subsection{Locality of the dressing transformation}

To prove Theorem \ref{thm: Thm2}, we will use the rapid decay of the $A_n$ together with repeated application the following lemma, which is very similar to Lemma \ref{lem: Exponential} but concerned with the different norm $\lVert \cdot \rVert_{\mu,x}$ that was introduced to describe operators with support near a site $x$.

\begin{lemma}\label{lem: LocOp}
Let $\mu' > \mu \geq \log 2$ and $A,B \in \caB$ with $\lVert A \rVert_{\mu'}\leq (\mu'-\mu)/2$, then
\begin{equation*}
\bigl\lVert \e^{\ad_A} B -B \bigr\rVert_{\mu,x} \leq \frac{2 \lVert A \rVert_{\mu'} \lVert B\rVert_{\mu',x}}{\mu' -\mu},
\end{equation*}
for all $x\in\Lambda$.
\end{lemma}
\noindent Let us write $U_n := \e^{-\i A_1}\dots\e^{-\i A_n}$, then we find, for every $O\in\caB$,
\begin{align*}
\bigl\lVert U^{-1}_n O U_n^{} \bigr\rVert_{\kappa,x}
& \leq \bigl\lVert \e^{-\i\ad_{A_n}}( U^{-1}_{n-1} O U_{n-1}^{})  \bigr\rVert_{2(n+1),x}\\
&\leq \Bigl(1+\frac{2 a_n}{\kappa_{2n}-\kappa_{2(n+1)}}\Bigr)\bigl\lVert U^{-1}_{n-1} O U_{n-1}^{}  \bigr\rVert_{2n,x}
\end{align*}
and repeating this step $n$ times gives
\begin{equation*}
\bigl\lVert U^{-1}_n O U_n^{} \bigr\rVert_{\kappa,x}\leq \lVert O\rVert_{\kappa',x} \prod_{m\geq 1} \Bigl(1+\frac{16 m^2 a_m}{\kappa'-\kappa}\Bigr)\leq \exp\Bigl(c\cdot\frac{ \lVert H'\rVert_{\kappa'}}{g(\kappa'-\kappa)}\Bigr)\lVert O\rVert_{\kappa',x},
\end{equation*}
which converges by the fast decay of the sequence $a_m$. The other locality estimate of the theorem, which is also meaningful for operators that may have full support throughout the entire volume, follows in the same way from Lemma \ref{lem: Exponential}.

\section{Proofs of the Lemmata}\label{sec: Proofs}

\subsection*{Proof of Lemma \ref{lem: Gap}}
We will compute the reduced resolvent $R(z)=\bar P (H_0+F-z)^{-1}$ perturbatively and show that its expansion converges for $\Re (z)\leq g/2$. First we introduce a partition of the identity $\opunit = \sum_{X\subset \Lambda} Q_X$ with projections $Q_X=\bar P_X P_{\Lambda\setminus X}$ and expand the resolvent in a Neumann series
\begin{equation}\label{eq: Neumann}
 R(z)=\sum_{\emptyset \neq X\subset \Lambda}R'(z) \sum_{n=0}^\infty \bigl(-F R'(z)\bigr)^n Q_{X}
\end{equation}
with a unperturbed reduced resolvent $R'(z)=\bar P (H_0-z)^{-1}$. Let us focus on the norm of the $n$-th term in the series and spell out $F$ as its sum of local terms,
\begin{align*}
&\bigl\lVert R'(z)\bigl(F R'(z)\bigr)^n Q_{X} \bigr\rVert\\
\leq&{\Bigl\lVert \sum_{\bsS_1,\dots,\bsS_n} R'(z)\prod_{i=1}^n F_{\bsS_i} Q_{X_i}  R'(z) \Bigr\rVert,}
\end{align*}
for uniquely defined non-empty $X_1,\dots,X_n\subset \Lambda$. {Here and in the following we use the order convention that the smallest index term in operator products is to the right. Each set $X_i$ only depends on $\bsS_1,\dots,\bsS_{i-1}$ or more precisely on $X_{i-1}$ and $\bsS_{i-1}$. Starting from $X = X_1$ the other sets $X_i$, $2\leq i\leq n$, are defined iteratively through $X_i=(X_{i-1}\setminus S_{i-1})\cup S_{i-1}^{\scriptscriptstyle -}$.} Note that $R'(z)Q_X=Q_XR'(z)$, $X\subset \Lambda$, and that we furthermore have the bound 
\begin{equation*}
{\lVert R'(z)Q_X\rVert\leq \frac{2}{g\lvert X\rvert},}
\end{equation*}
by the uniform gap condition on the $h_x$.
Finally we also use that $F_{\bsS}Q_X=0$ unless $S\cap X\neq \emptyset$, since $F$ is frustration-free. Therefore,
\begin{align*}
&\sum_{\bsS_1,\dots,\bsS_n}\Bigl\lVert R'(z)\prod_{i=1}^n F_{\bsS_i} Q_{X_i}  R'(z) \Bigr\rVert\\
\leq& \frac{2}{g}\sum_{\bsS_1,\dots,\bsS_{n-1}}\Bigl( \sum_{x\in X_n}\sum_{\bsS_n:\; x\in S_n}\frac{2 \lVert F_{\bsS_n}\rVert}{g \lvert X_n\rvert} \Bigr)\Bigl\lVert \prod_{i=1}^{n-1} F_{\bsS_i} Q_{X_i}  R'(z) \Bigr\rVert\\
\leq& \frac{2}{g} \Bigl(\frac{2\lVert F\rVert_{\mu=0}}{g}\Bigr)^n,
\end{align*}
so that the resolvent's expansion \eqref{eq: Neumann} is indeed bounded by a finite sum of convergent geometric series' if $\lVert F\rVert_{\mu=0}<g/2$. {By standard spectral perturbation theory, see e.g.\ \cite{Kat}, this finishes the proof of the lemma. In particular, the multiplicity of the gapped eigenvalue $0$ remains constant along the path $H_0+t F$, $t\in [0,1]$ and hence $\Omega$ is a non-degenerate eigenstate for $H_0+F$.} 

\subsection*{Proof of Lemma \ref{lem: PropG}}
By decomposing all operators as in \eqref{eq: Decomp}  we find that
\begin{align}\label{eq: PropG}
&\caV^\spl \bigl[(H_0+F_n) A_n\bigr]=-\i\caV^\spl \bigl[ (H_0+F_n)\caV^\spl[R_n V_n] \bigr]=-\i\caV^\spl\bigl[(H_0+F_n) R_n V_n\bigr]=-\i V^\spl_n.
\end{align}
Concerning the first step, note that, when inserted above, the first term in 
\begin{equation*}
A_n=\i\caV^ {\scriptscriptstyle -}[V_n R_n]-\i{\caV^ {\scriptscriptstyle +}}[R_n V_n]
\end{equation*}
does not contribute.
The inner $\caV^\spl$ operation can be dropped to get to the third equality, since either $\caV^\spl[(R_n V_n)_\bsS]=0$ or $\caV^\spl[(R_n V_n)_\bsS]=(R_n V_n)_\bsS$. And if it vanishes, then either $\bsS={\boldsymbol \emptyset}$ or $(H_0+F_n) (R_n V_n)_\bsS=O\bar P_x$, for some operator $O$ and site $x$, which is annihilated by the outer $\caV^\spl$ operation. Otherwise, concerning the possible constant term $ (R_n V_n)_{\boldsymbol \emptyset}$, note that 
\begin{equation}
\caV^{\spl}\bigl[(H_0+F_n)(R_nA_n)_{\boldsymbol{\emptyset}}\bigr]\propto\caV^{\spl}[H_0+F_n]=0,
\end{equation}
because $F_n$ and $H_0$ are frustration-free. In the same way, one also obtains $\caV^\smi[A_n(H_0+F_n)]= \i V_n^\smi$. Both equations together yield
\begin{equation}
\i\caV^\smi \bigl[A_n (H_0+F_n)\bigr]-\i\caV^\spl \bigl[(H_0+F_n) A_n\bigr]+V_n =\i\caV\bigl[[A_n,(H_0+F_n)]\bigr]+V_n=0,
\end{equation}
where it was again used that $H_0+F_n$ is frustration-free.

\subsection*{Proof of Lemma \ref{lem: Generator}}
First, we point out that it is sufficient to prove
\begin{equation*}
\bigl\lVert \caV^\spl [RV] \bigr\rVert_\mu \leq \frac{8 \lVert V^\spl \rVert_\mu}{g/4-\lVert F \rVert_\mu},
\end{equation*} 
under the general conditions of the lemma. Using this claim with $F^*$ and $V^*$ in the role of $F$ and $V$, we also find $\lVert V^{\smi}\rVert_\mu = \lVert (V^*)^{\spl}\rVert_\mu$ and hence
\begin{align*}
\bigl\lVert \caV^\smi [VR] \bigr\rVert_\mu=\bigl\lVert \bigl( \caV^\smi [VR]\bigr)^* \bigr\rVert_\mu &= \bigl\lVert \caV^\spl [R^*V^*] \bigr\rVert_\mu\\
&\leq\frac{8 \lVert V^\smi \rVert_\mu}{g/4-\lVert F \rVert_\mu},
\end{align*}
which would finish the proof of the lemma, because $\lVert V^\spl\rVert_\mu+\lVert V^\smi\rVert_\mu=\lVert V\rVert_\mu$.
We continue using the notation from the proof of Lemma \ref{lem: Gap} and set $R'=R'(z=0)$. Again, we expand the resolvent $R$ in $\caV^\spl[RV^\spl]$ as its Neumann series.
Note the following two properties of the $\caV$ operation. For all $O_\bsS\in\caB_\bsS$ and all $S_0\subset\Lambda$, we find that
\begin{equation*}
\caV^{\scriptscriptstyle +} \bigl[O_\bsS V_{S_0}^{\scriptscriptstyle +}\bigr]=0\qquad if\quad S^n\setminus S_0\neq\emptyset\quad \text{or}\quad S^{\scriptscriptstyle -}\setminus S_0\neq\emptyset
\end{equation*}
and $\caV^{\scriptscriptstyle +}[O_\bsS  P_{S'}]=\caV^{\scriptscriptstyle +}[O_\bsS]$ if $S\cap S'=\emptyset$.
Therefore
\begin{equation*}
\caV^\spl[O  V^\spl_{S_0}]=\caV^\spl[OQ_{S_0}V^\spl_{S_0}], \quad \text{for all } O\in \caB,
\end{equation*}
we can employ exactly the same convergent expansion \eqref{eq: Neumann} as in the previous proof to obtain
\begin{equation*}
\caV^\spl[RV^\spl_{S_0}]=\sum_{S}\caV^\spl[RV^\spl_{S_0}]^{}_{S}=\sum_{S}\caV^\spl\Bigl[ \sum_{k=0}^\infty \sum_{\bsS_1,\dots,\bsS_k}R'\Bigl(\prod_{i=1}^k F_{\bsS_i}Q_{X_i}R'\Bigr)Q_{S_0} V^\spl_{S_0} \Bigr]_{S}
\end{equation*}
For each $S_0$ and $\bsS_1,\dots,\bsS_k$ there is at most one $S=S(S_0,\bsS_1,\dots,\bsS_k)$, so that 
\begin{equation*}
\caV^\spl\Bigl[R'\Bigl(\prod_{i=1}^k F_{\bsS_i}Q_{X_i}R'\Bigr)Q_{S_0} V^\spl_{S_0} \Bigr]_{S}\neq 0.
\end{equation*}
The expression in between the brackets is of the form $O_S P_{\Lambda\setminus S}$, where $O_S\in\caB_S^\spl$, and by writing out $P_{\Lambda\setminus S}={\boldsymbol \otimes}_{x\in\Lambda\setminus S} (\opunit -(\opunit-P_x))$ it becomes clear that $\caV^\spl[O_SP_{\Lambda\setminus S} ]=O_S$.
Therefore, we find
\begin{equation*}
\Bigl\lVert\caV^\spl\Bigl[ R'\Bigl(\prod_{i=1}^k F_{\bsS_i}Q_{X_i}R'\Bigr)Q_{S_0} V^\spl_{S_0} \Bigr]_{S}\Bigr\rVert\leq \Bigl\lVert Q_S R'\Bigl(\prod_{i=1}^k F_{\bsS_i}Q_{X_i}R'\Bigr)Q_{S_0} V^\spl_{S_0} \Bigr\rVert
\end{equation*}
Since  $S\subset U_k:=\bigcup_{i=0}^k S_i$ and $w(U_k)\leq \sum_{i=0}^{k}w(S_i)$, we furthermore obtain
\begin{align*}
 &\bigl\lVert \caV^{\spl}[RV^{\spl}]\bigr\rVert_\mu\leq \frac{2}{g} \sum_{k=0}^\infty I_k\qquad \text{with}\\ 
 &I_k=\sup_x\sum_{S_0} \sum_{\bsS_1,\dots,\bsS_k} \chi(x\in U_k)\e^{\mu w(S_0)} \lVert V^\spl_{S_0}\rVert \prod_{i=1}^k \e^{\mu w(S_i)} \bigl\lVert Q_{X_{i+1}} F_{\bsS_i}R'Q_{X_i} \bigr\rVert
\end{align*}
where we set $X_{k+1}=S$ in each term, and $\chi$ denotes the indicator function. By induction in $k\geq 0$, we now show that
\begin{equation*}
I_k \leq  \Bigl(\frac{4\lVert F\rVert_\mu}{g}\Bigr)^k \lVert V^\spl \rVert_\mu,
\end{equation*}
which is obvious for $k=0$. For every $\bsS_k$, so that $F_{\bsS_k}\neq 0$, there is a site $z\in S_k$ satisfying $F_{\bsS_n} \bar P_z=F_{\bsS_k}$, because $F$ is frustration-free. We choose one such site $z =z(\bsS_k)$ for each $\bsS_k$. Using that each term in $I_k$ is zero unless $z\in U_{k-1}$, we get the induction step,
\begin{align*}
I_k\leq& \sup_x\sum_{S_0} \sum_{\bsS_1,\dots,\bsS_k} \bigl(\chi(x\in S_k)+\chi(x\in U_{k-1})\bigr)\e^{\mu w(S_0)} \lVert V^\spl_{S_0}\rVert \prod_{i=1}^k \e^{\mu w(S_i)} \bigl\lVert Q_{X_{i+1}} F_{\bsS_i}R'Q_{X_i} \bigr\rVert\\
\leq&\sup_x\sum_{S_0} \sum_{\bsS_1,\dots,\bsS_k} \chi(x\in S_k)\chi(z\in U_{k-1})\e^{\mu w(S_0)} \lVert V^\spl_{S_0}\rVert \prod_{i=1}^k \e^{\mu w(S_i)} \bigl\lVert Q_{X_{i+1}} F_{\bsS_i}R'Q_{X_i} \bigr\rVert\\
&{+\sup_x\sum_{S_0} \sum_{\bsS_1,\dots,\bsS_{k-1}} \chi(x\in U_{k-1}) \sum_{y\in X_k}\sum_{\bsS_k:\;y\in X_k}\frac{2 \lVert F_{\bsS_k}\rVert}{g\lvert X_k\rvert}\e^{\mu w(S_k)}} \\
&\hphantom{+\sup_x\sum_{S_0} \sum_{\bsS_1,\dots,\bsS_{k-1}}}{\times\e^{\mu w(S_0)} \lVert V^\spl_{S_0}\rVert \prod_{i=1}^{k-1} \e^{\mu w(S_i)} \bigl\lVert Q_{X_{i+1}} F_{\bsS_i}R'Q_{X_i} \bigr\rVert}\\
\leq&  \sup_x\sum_{\bsS_k:\; x\in S_k} \e^{\mu w(S_k)}\lVert F_{\bsS_k}R'\rVert\; I_{k-1}  + \frac{2\lVert F\rVert_\mu}{g} \; I_{k-1}\\
\leq & \;\frac{4\lVert F\rVert_\mu}{g} I_{k-1}
\end{align*}
which also finishes the proof of the lemma after explicitly computing the geometric series arising in the upper bound of $\sum_k I_k$.

\subsection*{Proof of Lemma \ref{lem: Commutator}}
First we write out the commutator using the locality of both operators
\begin{equation*}
[A,B]=\sum_{\bsS_1,\bsS_2: \; S_1\cap S_2\neq \emptyset} \bigl[A_{\bsS_1},B_{\bsS_2}\bigr]
\end{equation*}
and have a look at each local term $A_{\bsS_1}B_{\bsS_2}$.
For each pair $\bsS_1, \bsS_2$ there is another such index $\bsS'$ with support $S'\subset (S_1\cup S_2)\setminus (S_1^\smi\cap S_2^\spl)$ and an operator $O_{\bsS'}\in\caB_{\bsS'}$ with norm bounded by $\lVert O_{\bsS'}\rVert\leq \lVert A_{\bsS_1}\rVert \lVert B_{\bsS_2}\rVert$, such that $A_{\bsS_1}B_{\bsS_2}=P_{S_1^\smi\cap S_2^\spl}\otimes O_{\bsS'}$. Writing out again the ground state projection at each site $x\in S_1^\smi\cap S_2^\spl$ as $P_x = \opunit -(\opunit - P_x)$, and since 
\begin{equation*}
w(S_1\cup S_2)+\lvert S_1^\smi\cap S_2^\spl\rvert\leq w(S_1)+w(S_2)
\end{equation*}
and $\mu >\log 2$ by assumption,  we find
\begin{equation*}
\bigl\lVert A_{\bsS_1}B_{\bsS_2} \bigr \rVert_\mu\leq 2^{S_1\cap S_2} \lVert A_{\bsS_1} \rVert \lVert B_{\bsS_2} \rVert\leq \e^{\mu w(S_1)}\lVert A_{\bsS_1} \rVert \e^{\mu w(S_2)}\lVert B_{\bsS_2} \rVert.
\end{equation*}
We then get that the second term in $\lVert [A,B] \rVert_\mu=\lVert [A,B]_{\boldsymbol \emptyset} \rVert/\lvert \Lambda\rvert+\lVert [A,B] \rVert_\mu'$ is bounded above by
\begin{align*}
\bigl\lVert [A,B] \bigr\rVert_\mu' &\leq \sup_x \sum_{\bsS_1:\; x\in S_1} \sum_{\bsS_2:\; S_1\cap S_2\neq\emptyset}\Bigl(\bigl\lVert A_{\bsS_1}B_{\bsS_2} \bigr \rVert_\mu +\bigl\lVert A_{\bsS_2}B_{\bsS_1} \bigr \rVert_\mu +\bigl\lVert B_{\bsS_2} A_{\bsS_1} \bigr \rVert_\mu +\bigl\lVert B_{\bsS_1} A_{\bsS_2} \bigr \rVert_\mu\Bigr)\\
&\leq 2\sup_x \sum_{\bsS_1:\; x\in S_1}\sum_{y\in S_1}\sum_{\bsS_2:\; y\in S_2}\e^{\mu w(S_1)+\mu w(S_2)}\Bigl(\lVert A_{\bsS_1} \rVert \lVert B_{\bsS_2} \rVert +\lVert A_{\bsS_2} \rVert \lVert B_{\bsS_1} \rVert \Bigr)\\
&\leq 2\sup_x \sum_{\bsS_1:\; x\in S_1} \lvert S_1 \rvert \e^{\mu w(S_1)} \Bigl(\lVert A_{\bsS_1} \rVert \lVert B \rVert_\mu +\lVert A \rVert_\mu \lVert B_{\bsS_1} \rVert \Bigr)\\
& \leq \frac{4 \lVert A \rVert_{\mu'} \lVert B\rVert_{\mu'}}{\mu'-\mu},
\end{align*}
where we used $\lvert S_1\rvert\e^{-(\mu'-\mu)w(S_1)}\leq 1/(\mu'-\mu)$ for the last step. Concerning the first term, using
\begin{equation*}
\bigl\lVert \bigl( A_{\bsS_1}B_{\bsS_2}\bigr)_{\boldsymbol \emptyset} \bigr \rVert \leq\lVert A_{\bsS_1} \rVert\lVert B_{\bsS_2} \rVert,
\end{equation*}
we similarly obtain
\begin{align*}
\bigl\lVert [A,B]_{\boldsymbol \emptyset} \bigr\rVert&\leq  \sum_{x\in\Lambda}  \sum_{\bsS_1:\; x\in S_1} \sum_{\bsS_2:\; S_1\cap S_2\neq \emptyset}\Bigl(\bigl\lVert \bigl( A_{\bsS_1}B_{\bsS_2}\bigr)_{\boldsymbol \emptyset} \bigr \rVert +\bigl\lVert \bigl(A_{\bsS_2}B_{\bsS_1}\bigr)_{\boldsymbol \emptyset} \bigr \rVert +\bigl\lVert \bigl( B_{\bsS_2} A_{\bsS_1}\bigr)_{\boldsymbol \emptyset} \bigr \rVert +\bigl\lVert\bigl( B_{\bsS_1} A_{\bsS_2}\bigr)_{\boldsymbol \emptyset} \bigr \rVert\Bigr)\\
& \leq \lvert \Lambda \rvert \cdot 2\sup_x \sum_{\bsS_1:\; x\in S_1}\sum_{y\in S_1}\sum_{\bsS_2:\; y\in S_2}\Bigl(\lVert A_{\bsS_1} \rVert \lVert B_{\bsS_2} \rVert +\lVert A_{\bsS_2} \rVert \lVert B_{\bsS_1} \rVert \Bigr)\\
&\leq\lvert \Lambda \rvert \cdot 2\sup_x \sum_{\bsS_1:\; x\in S_1} \lvert S_1 \rvert  \Bigl(\lVert A_{\bsS_1} \rVert \lVert B \rVert_\mu +\lVert A \rVert_\mu \lVert B_{\bsS_1} \rVert \Bigr)\\
& \leq \lvert \Lambda \rvert \cdot\frac{4 \lVert A \rVert_{\mu'} \lVert B\rVert_{\mu'}}{\mu'-\mu},
\end{align*}
which proves the lemma.

\subsection*{Proof of Lemma \ref{lem: Exponential}}
This lemma and its proof is essentially the same as Lemma 4.1 in \cite{AbaDeRHuvW} with adaptations due to the different type of norm (there, on potentials) employed in this reference. The proof is based on a combinatorial trick to reduce graph structure inductively, which is commonly used for controlling cluster expansions in general polymer models, see \cite{Uel}. Before that, we expand equation \eqref{eq: BoundExp} into local terms,
\begin{equation*}
\sum_{k=1}^{\infty}\frac{\ad_A^k B^{\scriptscriptstyle (k)} }{k!}=\sum_{k=1}^{\infty} \frac{1}{n!}\sum_{\bsS_0,\dots,\bsS_k}\ad_{A_{\bsS_k}} \dots \ad_{A_{\bsS_1}} B_{\bsS_0}^{\scriptscriptstyle (k)}.
\end{equation*}
Generalizing only slightly the argument in the previous proof for the estimate concerning the product $A_{\bsS_1}B_{\bsS_2}$, one can rewrite a product of several operators $O^{\scriptscriptstyle (i)}_{\bsS_i}\in\caB_{\bsS_i}$, $i=0, \dots, k$, as
\begin{equation*}
O^{\scriptscriptstyle (k)}_{\bsS_k}\dots O^{\scriptscriptstyle (0)}_{\bsS_0}= P_{S''}\otimes O_{\bsS'}, \qquad O_{\bsS'}\in\caB_{\bsS'},
\end{equation*}
where $S'$ and $S''$ are disjoint and satisfy
\begin{equation*}
S'\cup S''=U_k:={\textstyle \bigcup}_{i=0}^k S_i\quad \text{and}\quad S''\subset {D_k := {\textstyle \bigcup}_{0\leq i<j\leq k} (S_{i}\cap S_j).}
\end{equation*}
Moreover, the norm of $O_{\bsS'}$ is bounded by $\lVert O_{\bsS'}\rVert\leq \prod_i \lVert O_{\bsS_i}^{\scriptscriptstyle (i)}\rVert$.
Therefore, for every $\bsS_0,\dots, \bsS_k$,
\begin{align}\label{eq: BoundAds}
\sum_{\bsS}\Bigl\lVert \bigl(\ad_{A_{\bsS_k}} \dots \ad_{A_{\bsS_1}} B_{\bsS_0}^{\scriptscriptstyle (k)}  \bigr)_{\bsS} \Bigr\rVert
\leq {2^{ \lvert D_k \rvert}} \lVert B_{\bsS_0}^{\scriptscriptstyle (k)}\rVert \prod_{j=1}^{k}2 \lVert A_{\bsS_j}\rVert,
\end{align}
which moreover vanishes whenever one of the $S_j$ does not intersect $U_{j-1}$. Clearly $S$ is either empty or $S\subset U_k$ for each term in the sum. Moreover, the expression is zero if one of the $S_i$ equals the empty set or if the sets from the sequence $S_0,\dots,S_k$ can be divided into two (non-empty) families of mutually disjoint sets. If such a partition is not possible, the sequence is called a \emph{cluster}. {If $S_0,\dots,S_k$ is a cluster then we have}
\begin{equation}\label{eq: BoundWeight}
{\mu w(S)+\log 2\cdot  \lvert D_k \rvert\leq\mu w(U_k)+\log 2 \cdot \lvert D_k \rvert\leq \mu \sum_{i=0}^k w(S_i)}.
\end{equation}
{ Recall the assumption $\mu\geq \log 2$, then the last inequality follows from
\begin{equation*}
\begin{split}
& \lvert U_k\rvert + \lvert D_k\rvert\leq \sum_{i=0}^k \lvert S_i\rvert\qquad \text{and}\\
& \lvert U_k\rvert_c = \min\bigl\{\lvert T\rvert \mid T \subset\bbZ^v \text{ connected},\, U_k\subset T\bigr\}\leq \sum_{i=0}^k \lvert S_i\rvert_c
\end{split}
\end{equation*}
where we used that $S_0,\dots,S_k$ is a cluster for the second inequality (hence the union of the minimal connected extensions of the $S_i$ is a connected set with not less than $\lvert U_k\rvert_c$ elements).
} Therefore we arrive at
\begin{align*}
\frac{\bigl\lVert \ad^k_A B^{\scriptscriptstyle (k)} \bigr\rVert_\mu'}{k!} &\leq \sup_x \frac{1}{k!} \sum_{\bsS:\; x\in S} \e^{\mu w(S)}\sum_{S_0,\dots,S_k}^{\mathrm{cluster}} \sum_{\substack{\bsS_0,\dots,\bsS_k:\\ S(\bsS_i)=S_i}}\Bigl\lVert \bigl(\ad_{A_{\bsS_n}} \dots \ad_{A_{\bsS_1}} B_{\bsS_0}^{\scriptscriptstyle (k)}  \bigr)_{\bsS} \Bigr\rVert\\
&\leq \sup_x \frac{1}{k!}  \sum_{\substack{S_0,\dots,S_k:\\x\in U_k}}^{\mathrm{cluster}}\e^{\mu w(U_k)}2^{ \lvert D_k \rvert} \sum_{\substack{\bsS_0:\\ S(\bsS_0)=S_0}}\lVert B_{\bsS_0}^{\scriptscriptstyle (k)}\rVert\prod_{j=1}^k \Bigl(\sum_{\substack{\bsS_j:\\ S(\bsS_j)=S_j}}2\lVert A_{\bsS_j}\rVert\Bigr)\\
&\leq \frac{18 b \lVert A\rVert_{\mu'} }{(\mu'-\mu)^2}\sup_x \frac{1}{(k+1)!} \sum_{\substack{S_0,\dots,S_k:\\x\in U_k}}^{\mathrm{cluster}} \prod_{i=0}^k v(S_i),
\end{align*}
where we made use of the assumption $\lVert A\rVert_\mu\leq (\mu'-\mu)/6$, and we also introduced
\begin{equation*}
v(S)={(\mu'-\mu)} \sum_{\bsS:\;S(\bsS)=S}\frac{\e^{\mu w(S)}}{3}\Bigl(\frac{2\lVert A_\bsS\rVert}{\lVert A\rVert_{\mu'}}+\frac{\lVert B_\bsS\rVert}{b}\Bigr).
\end{equation*}
In a similar but more lavish manner, we can show that
\begin{align*}
\frac{\bigl\lVert\bigl( \ad^k_A B^{\scriptscriptstyle (k)}\bigr)_{\boldsymbol \emptyset} \bigr\rVert}{k!}&\leq  \frac{1}{k!} \sum_{S_0,\dots,S_k}^{\mathrm{cluster}} \sum_{\substack{\bsS_0,\dots,\bsS_k:\\ S(\bsS_i)=S_i}}\Bigl\lVert \bigl(\ad_{A_{\bsS_n}} \dots \ad_{A_{\bsS_1}} B_{\bsS_0}^{\scriptscriptstyle (k)}  \bigr)_{\boldsymbol \emptyset} \Bigr\rVert\\
&\leq \lvert \Lambda \rvert \cdot\frac{18 b \lVert A\rVert_{\mu'} }{(\mu'-\mu)^2}\sup_x \frac{1}{(k+1)!} \sum_{\substack{S_0,\dots,S_k:\\x\in U_k}}^{\mathrm{cluster}} \prod_{i=0}^k v(S_i).
\end{align*}
Next, we use induction (in $N$) to show that, for every subset $S'$,
\begin{equation}\label{eq: Cluster}
1+\sum_{k=0}^N \frac{1}{(k+1)!} \sum_{S_0,\dots,S_k} \chi\bigl({\textstyle{ S_0,\dots,S_k,S' \atop  \text{ is a cluster}}}\bigr)\prod_{i=0}^{k} v(S_i)\leq \e^{(\mu'-\mu)w(S')}.
\end{equation}
The following statement again holds for every $S'$ and is in fact stronger than required for starting the induction at $N=0$: 
\begin{equation}\label{eq: KoteckyPreis}
\sum_{S:\; S\cap S'\neq \emptyset}v(S)\e^{(\mu'-\mu)w(S)}\leq \lvert S'\rvert \sup_x \sum_{S\ni x} v(S)\e^{(\mu'-\mu)w(S)}\leq (\mu'-\mu) w(S').
\end{equation}
For general $N$, we reorganize \eqref{eq: Cluster} and collect terms in which at least $m$ of the sets $S_0,\dots,S_k$ intersect with $S'$. Every such set $S''$ that has overlap with $S'$ can be part of a cluster with at most $N$ other sets from $S_0,\dots,S_k$, so that the induction hypothesis can be used. Therefore we obtain the upper bound
\begin{align*}
& 1+\sum_{k=0}^N \frac{1}{(k+1)!} \sum_{S_0,\dots,S_k} \chi\bigl({\textstyle{ S_0,\dots,S_k,S' \atop  \text{ is a cluster}}}\bigr)\prod_{i=0}^{k} v(S_i)\\
\leq\, & 1+ \sum_{m=1}^{N}\frac{1}{m!}\Bigl[ \sum_{S'':\; S''\cap S'\neq \emptyset} v(S'')\Bigl(1+\sum_{M=0}^{N-1}\frac{1}{(M+1)!} \sum_{S_0,\dots,S_M} \chi\bigl({\textstyle{ S_0,\dots,S_M,S'' \atop  \text{ is a cluster}}}\bigr)\prod_{i=1}^M v(S_i)\Bigr) \Bigr]^m\\
\leq \, & 1+ \sum_{m=1}^{N}\frac{1}{m!}\Bigl[ \sum_{S'':\; S''\cap S'\neq \emptyset} v(S'')\e^{(\mu'-\mu)w(S'')} \Bigr]^m,
\end{align*}
and we can use \eqref{eq: KoteckyPreis} to conclude that \eqref{eq: Cluster} holds indeed for all $N\geq 0$. In particular, we may evaluate  equation \eqref{eq: Cluster} for any singleton $S'=\{x\}$. Finally, recall the assumption $\mu'-\mu\leq 1$, so that $\e^{(\mu'-\mu)w(\{x\})}-1\leq7 (\mu'-\mu)$, and our estimates can be assembled in the following way to prove the lemma,
\begin{align*}
\Bigl\lVert \sum_{k=1}^{\infty}\frac{\ad_A^k B^{\scriptscriptstyle (k)} }{k!} \Bigr\rVert_\mu& \leq \sum_{k=1}^{\infty} \frac{1}{k!}\Bigl(\bigl\lVert\bigl( \ad^k_A B^{\scriptscriptstyle (k)}\bigr)_{\boldsymbol \emptyset} \bigr\rVert\bigr/\lvert \Lambda\rvert+ \bigl\lVert \ad^k_A B^{\scriptscriptstyle (k)} \bigr\rVert_\mu' \Bigr)\\
&\leq 2 \cdot \frac{18 b \lVert A\rVert_{\mu'} }{(\mu'-\mu)^2}\sum_{k=1}^\infty \sup_x \frac{1}{(k+1)!} \sum_{S_0,\dots,S_k} \chi\bigl({\textstyle {S_0,\dots,S_k,\{x\} \atop \text{is a cluster}}}\bigr)\prod_{i=0}^nv(S_i)\\
&\leq 2\cdot\frac{7\cdot18 b \lVert A\rVert_{\mu'}}{\mu'-\mu}.
\end{align*}

\subsection*{Proof of Lemma \ref{lem: LocOp}}

We proceed very similarly and with the same notation as in the proof of Lemma \ref{lem: Exponential}. We expand the exponential and use again \eqref{eq: BoundAds}, but this time with
\begin{equation*}
\mu w_x(S)+\log 2  \sum_{j}\lvert S_{j-1}\cap S_{j} \rvert\leq \mu w_x(S_0) + \mu\sum_j w(S_j) 
\end{equation*}
instead of \eqref{eq: BoundWeight}. We obtain
\begin{align*}
\frac{\bigl\lVert \ad_A^k B\bigr\rVert_{\mu,x}}{k!}&\leq \sum_{\bsS} \e^{\mu w_x(S)}\frac{1}{k!} \sum_{S_0, \dots,S_k}^{\mathrm{cluster}} \sum_{\substack{\bsS_0,\dots,\bsS_k:\\ S(\bsS_i)=S_i}}\Bigl\lVert \bigl(\ad_{A_{\bsS_n}} \dots \ad_{A_{\bsS_1}} B_{\bsS_0}^{\scriptscriptstyle (k)}  \bigr)_{\bsS} \Bigr\rVert\\
&\leq \sum_{\bsS_0} \e^{\mu w_x(S_0)} \lVert B_{\bsS_0}\rVert \frac{1}{k!} \sum_{S_1,\dots,S_k} \chi\bigl({\textstyle{ S_0,\dots,S_k \atop  \text{ is a cluster}}} \bigr)\prod_j \e^{\mu w(S_j)}\Bigl(\sum_{\substack{\bsS_j:\\ S(\bsS_j)=S_j}}2\lVert A_{\bsS_j}\rVert\Bigr)\\
&\leq \frac{2 \lVert A\rVert_{\mu'}}{\mu'-\mu}\sum_{\bsS_0} \e^{\mu w_x(S_0)} \lVert B_{\bsS_0}\rVert \frac{1}{k!} \sum_{S_1,\dots,S_k} \chi\bigl({\textstyle{ S_0,\dots,S_k \atop  \text{ is a cluster}}} \bigr)\prod_j \widetilde{v}(S_j),
\end{align*}
using the assumption $\lVert A\rVert_{\mu'}\leq (\mu'-\mu)/2$ in the last step, and where we introduced a different (weight) function 
\begin{equation*}
\widetilde{v}(S) := \frac{(\mu'-\mu)}{\lVert A \rVert_{\mu'}} \sum_{\bsS:\;S(\bsS)=S} \e^{\mu w(S)} \lVert A_{\bsS} \rVert.
\end{equation*}
Just as we got to \eqref{eq: Cluster}, we find 
\begin{equation*}
\sum_{k=1}^\infty \frac{1}{k!} \sum_{S_1,\dots,S_k} \chi\bigl({\textstyle{ S_0,\dots,S_k \atop  \text{ is a cluster}}} \bigr)\prod_j \widetilde{v}(S_j)\leq \e^{(\mu' -\mu) w(S_0)}
\end{equation*}
which finishes the proof if insert it above and sum over $k$ while keeping in mind that
\begin{equation*}
\e^{\mu w_x(S_0)}\e^{(\mu'-\mu) w(S_0)}\leq \e^{\mu' w_x(S_0)}.
\end{equation*}


\begin{thebibliography}{99}


\bibitem{AbaDeRHuvW} D.~Abanin, W.~De Roeck, F.~Huveneers, and W.~W.~Ho, \emph{Asymptotic energy conservation in periodically driven many-body systems}, arXiv:1509.05386 (2015)

\bibitem{Alb} C.~Albanese, \emph{Unitary dressing transformations and exponential decay below threshold
for quantum spin systems}, Commun.~Math.~Phys.~\textbf{134}, 1-27, 237-272 (1990)

\bibitem{Bachmann} S.~Bachmann, S.~Michalakis, B.~Nachtergaele, and R.~Sims, \emph{Automorphic equivalence within gapped phases of quantum lattice systems}, Commun.~Math.~Phys.~\textbf{309}, 835-871 (2011)

\bibitem{BraCubLucMicPer} F.\ Brandao, T.~Cubitt, A.~Lucia , S.~Michalakis, and D.~Perez-Garcia, \emph{ Area law for fixed points of rapidly mixing dissipative quantum systems}, J.\ Math.\ Phys.\ \textbf{56}, 102202 (2015).

\bibitem{BraHas} S.~Bravyi and M.~Hastings, \emph{A short proof of stability of topological order under local perturbations}, Commun.~Math.~Phys.~\textbf{307}, 609-627 (2011).

\bibitem{CraDerSch}  N.~Crawford, W.~De Roeck, and M.~Sch\"utz, \emph{Uniqueness regime for Markov dynamics on quantum lattice spin systems}, J.~Phys.~A:~Math.~Theor.~\textbf{48}, 425203 (2015).

\bibitem{CubLucMicPer} T.~Cubitt, A.~Lucia , S.~Michalakis, and D.~Perez-Garcia, \emph{Stability of local quantum dissipative systems}, Commun.~Math.~Phys.~\textbf{337}, 1275-1315 (2011).

\bibitem{DatKen} N.~Datta and T.~Kennedy, \emph{Expansions for one quasiparticle states in spin 1/2 systems}, J.~Stat.~Phys.~\textbf{108}, 373-399 (2002)

\bibitem{DerSch} W.~De Roeck and M.~Sch\"utz, \emph{Local perturbations perturb -exponentially- locally}, J.~Math.Phys.~\textbf{56}, 061901 (2015)


\bibitem{DerMaeNetSch}  W.~De Roeck, C.~Maes, K.~Neto\v{c}n\'y, and M.~Sch\"utz, \emph{Locality and nonlocality of classical restrictions of quantum spin systems with applications to quantum large deviations and entanglement}, J.~Math.~Phys.~\textbf{56}, 023301 (2015).

\bibitem{FanNacWer} M.\ Fannes, B.\ Nachtergaele, and R.\ Werner, \emph{Finitely Correlated States on
Quantum Spin Chains}, Comm.\ Math.\ Phys.\ \textbf{144}, 443-490 (1992).

\bibitem{Has} M.~Hastings, \emph{Lieb--Schultz--Mattis in higher dimensions}, Phys.~Rev.~B \textbf{69}, 104431 (2004).

\bibitem{Has2} M.~Hastings, \emph{Solving gapped Hamiltonians locally}, Phys.\ Rev.\ B \textbf{73}, 085115 (2006).

\bibitem{KasTem} M.\ Kastoryano and K.\ Temme, \emph{Quantum logarithmic Sobolev inequalities and rapid mixing}, J.\ Math.\ Phys.\ \textbf{54}, 052202 (2013).

\bibitem{Kat} T.\ Kato, \emph{Perturbation theory of linear operators}, Springer-Verlag (1995).

\bibitem{KenTas} T.~Kennedy and H.~Tasaki, \emph{Hidden symmetry breaking and the Haldane phase in $S = 1$ quantum spin chains}, Commun.~Math.~Phys.~\textbf{147}, 431–484 (1992)

\bibitem{KirTho} J.~R.~Kirkwood and L.~E.~Thomas, \emph{Expansions and phase transitions for the ground states of quantum Ising lattice
systems}, Commun.~Math.~Phys.~\textbf{88}, 569-580 (1983).

\bibitem{KraEtAl} B.\ Kraus, H.\ Büchler, S.\ Diehl, A.\ Kantian, A.\ Micheli, and P.\ Zoller, \emph{Preparation of entangled states by quantum Markov processes}, Phys.\ Rev.\ A \textbf{78}, 042307 (2008).

\bibitem{MicZwo} S.~Michalakis and J.~Zwolak, \emph{Stability of frustration-free Hamiltonians}, Commun. Math.~Phys.~\textbf{322}, 277-302 (2011).

\bibitem{NacOgaSim} B.\ Nachtergaele, Y.\ Ogata, and R.\ Sims, \emph{Propagation of correlations in quantum lattice systems}, J.\ Stat.\ Phys.\ \textbf{124}, 1-13 (2006).

\bibitem{Nac} B.\ Nachtergaele, \emph{The spectral gap for some spin chains with discrete symmetry
breaking}. Commun.\ Math.\ Phys.\ \textbf{175}, 565-606 (1996).

\bibitem{NacVerZag} B.\ Nachtergaele, A.\ Vershynina, and V.\ Zagrebnov, \emph{Lieb-Robinson bounds and existence of the thermodynamic limit for a class of irreversible quantum dynamics}, AMS Contemp.\ Math.\ \textbf{552}, 161-175(2011).

\bibitem{NetRed} K.~Neto\v{c}n\'y and F.~Redig, \emph{Large deviations for quantum spin systems}, J.~Stat.~Phys.~\textbf{117}, 521-547 (2004)

\bibitem{Pou} D.\ Poulin, \emph{Lieb-Robinson bound and locality for general Markovian quantum dynamics},  	Phys.\ Rev.\ Lett.\ \textbf{104}, 190401 (2010).

\bibitem{SpiSta} W.\ Spitzer and S.\ Starr, \emph{Improved bounds on the spectral gap above frustration-
free ground states of quantum spin chains}. Lett.\ Math.\ Phys.\ \textbf{63}, 165-177
(2003).

\bibitem{SzeWol} O.\ Szehr and M.\ Wolf, \emph{Perturbation theory for parent Hamiltonians of matrix
product states}. J.\ Stat.\ Phys.\ \textbf{159}, 752-771 (2015).

\bibitem{VerWolCir} F.\ Verstraete, M.\ Wolf, and J.\ Cirac. \emph{Quantum computation and quantum-state engineering
driven by dissipation}, Nat.\ Phys.\ \textbf{5}, 633-636 (2009).

\bibitem{Uel} D.~Ueltschi, \emph{Cluster expansions and correlation functions}, Mosc.~Math.~J.~\textbf{4}, 511-522 (2004)

\bibitem{Yar04} D.~A.~Yarotsky, \emph{Perturbations of ground states in weakly interacting quantum spin systems}, J.~Math.~Phys.~\textbf{45}, 2134 (2004).

\bibitem{Yar05} D.~A.~Yarotsky, \emph{Uniqueness of the ground state in weak perturbations of non-interacting gapped quantum lattice Systems}, J.~Stat.~Phys.~\textbf{118}, 119-144 (2005).

\bibitem{Yar06} D.~A.~Yarotsky, \emph{Ground states in relatively bounded quantum perturbations of classical lattice systems}, Commun.~Math.~Phys.~\textbf{261}, 569-580 (1983).

\bibitem{Zni} M.\ Znidaric, \emph{Relaxation times of dissipative many-body quantum systems}, Phys.\ Rev.\ E \textbf{92}, 042143 (2015).




 




 
\end{thebibliography}
\end{document}